The tex document is followed by the postscript file for figure 1, all text
after the \end{document} line is postscript.

\documentstyle[preprint,revtex]{aps}
\begin{document}
 \draft

\begin{title}
Angular Momentum Analysis of Spin $\frac{1}{2} \times \frac{1}{2}$ Scattering\\
including Singlet-Triplet Mixing
\end{title}

        \author{T. Mefford and R.H. Landau}
\begin{instit}
    Physics Department, Oregon State University, Corvallis, OR 97331
\end{instit}

\receipt{\today}

\begin{abstract}
The angular momentum (partial wave) reduction of the
Lippmann--Schwinger equation describing the interaction of two spin
$\frac{1}{2}$ particles is extended to the case in which the spin
singlet and triplet states are coupled. A straight forward method for
obtaining the angular momentum decomposition of the general potential
is indicated. The derived formalism is needed to describe the
interaction between two nonidentical spin--$\frac{1}{2}$ particles or
between two nucleons when isospin symmetry is violated.  The resulting
modification of the Stapp phase--shift analysis is given.\\
\end{abstract}


\narrowtext

\section{INTRODUCTION}

Recent experimental advances are permitting detailed measurements and
analyses of the scattering of polarized nucleons from polarized
nuclei\cite{ray}. This, in turn, is leading to a rigorous test of
the theory describing spin-spin scattering\cite{otto}. The general form of the
elastic scattering $T$ matrix for two spin $\frac{1}{2}$ particles can
be quite complicated.  It simplifies considerably if we assume
rotation invariance, parity conservation, and time reversal
invariance, in which case\cite{laf,bys,gold,csonka,ger}:

\begin{eqnarray}
    2T(\vec{k'},\vec{k}) &=& a(\vec{k'},\vec{k})+b(\vec{k'},\vec{k})
+ (a-b)\vec{\sigma}^{p}_{n} \vec {\sigma}^{t}_{n} +
\left(c(\vec{k'},\vec{k})+d(\vec{k'},\vec{k})\right) \vec{\sigma}^{p}_{m}
\vec{\sigma}^{t}_{m} \nonumber\\ & &+ (c-d) \vec{\sigma}^{p}_{l}
\vec{\sigma}^{t}_{l} + e(\vec{k'},\vec{k}) ( \vec{\sigma}^{p}_{n} +
\vec{\sigma}^{t}_{n}) + f(\vec{k'},\vec{k}) ( \vec{\sigma}^{p}_{n} -
\vec{\sigma}^{t}_{n}) \label{T}
\end{eqnarray}
Here $\vec{k}$ and $\vec{k'}$ are the initial and final momenta in the
center--of--mass system, $\vec{\sigma^{p}}$ and $\vec{\sigma^{t}}$ are
the Pauli spin operators for the projectile and target respectively,
and the subscript on each $\vec{\sigma}$ indicates a dot products with
one of the three independent unit vectors:
\begin{equation}
   \hat{n} = \frac{\vec{k} \times \vec{k'}} {|\vec{k} \times \vec{k'}|},\ \ \
   \hat{m} = \frac{\vec{k'} - \vec{k}} {|\vec{k'} - \vec{k}|},\ \ \
   \hat{l} = \frac{\vec{k} + \vec{k'}} {|\vec{k} + \vec{k'}|}   \label{vectors}
\end{equation}
For example, $\vec{\sigma}^{t}_{m} = \vec{\sigma}^{t} \cdot \hat{m}$.
For experimental (on-energy-shell) scattering, the $a-f$ coefficients
in (\ref{T}) are complex functions of the energy and scattering angle.
Once these coefficients are known, all 36 possible experimental spin
observables can be calculated from them\cite{laf,bys}.

Just as (\ref{T}) is the most general scattering amplitude expressed
as an operator in the direct product spin spaces of the projectile
and target, so the projectile-target potential must have the same
form:
\begin{eqnarray}
2V(\vec{k'},\vec{k}) &=& V_{a+b}(\vec{k'},\vec{k}) +
V_{a-b}(\vec{k'},\vec{k})\vec{\sigma}^{p}_{n}\vec{\sigma}^{t}_{n} +
V_{c+d}(\vec{k'},\vec{k})\vec{\sigma}^{p}_{m}\vec{\sigma}^{t}_{m}
\nonumber \\
	& & +
V_{c-d}(\vec{k'},\vec{k})\vec{\sigma}^{p}_{l}\vec{\sigma}^{t}_{l} +
V_{e}(\vec{k'},\vec{k})(\vec{\sigma}^{p}_{n} + \vec{\sigma}^{t}_{n}) +
V_{f}(\vec{k'},\vec{k})(\vec{\sigma}^{p}_{n} - \vec{\sigma}^{t}_{n}) \label{V}
\end{eqnarray}
where the subscripts on $V$ indicate the correspondence with
(\ref{T}).  If the potential has the structure (\ref{V}), then its use in a
wave equation will generate a scattering amplitude of the form
(\ref{T}). In traditional language, the $V_{a+b}$ term is a central
potential, the $V_{a-b}, V_{c+d},$ and $V_{c-d}$ terms are tensor
forces (dipole-dipole interactions for two electrons), the $V_{e}$
term is the usual spin-orbit potential\cite{gold}, and the $V_{f}$ term is
the unusual spin-orbit  potential which couples spin singlet and triplet
states.

Prior angular momentum (partial wave) analyses, such as those found in
Goldberger and Watson\cite{gold}, Stapp\cite{stapp} and
Goddard\cite{god}, have examined the scattering of two
spin--$\frac{1}{2}$ particles for the nucleon-nucleon case where the
$f$ amplitude vanishes due to the isospin symmetry. In our work on
proton scattering from spin--$\frac{1}{2}$ nuclei, we have had to
develop a Stapp--like partial wave analysis\cite{wave} for cases in
which the $f$ term does not vanish (extensions of the analysis of
observables including an $f$ amplitude had already been given by La
France and Winternitz\cite{laf}). While the work of Gersten\cite{ger}
provides a partial wave analysis, it is in the helicity---not angular
momentum--- representation. While we suspect that others have deduced
the general angular momentum analysis in the course of their research,
we have been unable to find those analyses in the literature and so wish
to present ours. Explicitly, we shall derive the partial--wave
decomposition of the Lippmann--Schwinger equation when there is mixing
between the spin singlet and triplet states (as well as within the
triplet), and indicate how we have applied the method in our
calculations.

\section{ANGULAR MOMENTUM ANALYSIS With SINGLET-TRIPLET MIXING}

Many--body effects and relativity leads to a potential $V$
incorporating complicated nonlocalities. For this reason, rather than
solve a Schr\"odinger equation we solve the equivalent, momentum space
Lippmann--Schwinger equation:

\begin{equation}
 T(\vec{k}',\vec{k}) = V(\vec{k}',\vec{k}) + \int \frac{d^{3}p}{
 E^{+} - E(p)} \, V(\vec{k}',\vec{p}) \, T(\vec{p},\vec{k}) \label{LS}
\end{equation}
where $E(p)=E_{p}(p)+E_{t}(p)$ is the projectile plus target energy
and the $+$ superscript indicates a positive $i\epsilon$ has been
added to the on--shell energy $E=E_{p}(k_{0})+E_{t}(k_{0})$. Since the
solution of a three--dimensional integral equation is a bit too
formidable, we reduce it to coupled one--dimensional equations by
expanding $T$ and $V$ in partial waves\cite{wave}:
\begin{eqnarray}
V(\vec{k'},\vec{k}) &=& \frac{2}{\pi} \sum_{jm_{j}ll'ss'} i^{(l' - l)}
\,V^{j(s's)}_{l'l} (k',k) \,{\cal Y}^{jm_{j}}_{l's'}(\hat{k}')
\,{\cal Y}_{ls}^{\dagger jm_{j}}(\hat{k}) \label{expand}\\
T(\vec{k'},\vec{k}) &=& \frac{2}{\pi} \sum_{jm_{j}ll'ss'} i^{(l' - l)}
\,T^{j(s's)}_{l'l} (k',k)
\,{\cal Y}^{jm_{j}}_{l's'}(\hat{k'}) \,{\cal Y}_{ls}^{\dagger jm_{j}}(\hat{k})
\label{expand2} \end{eqnarray}
Here $l$ and $s$ are the total orbital angular momentum and  total spin for
the target plus projectile, and $j$ is the total angular momentum:
\begin{eqnarray}
\vec{j} &=& \vec{l} + \vec{s}\\
 \vec{s} &=& \frac{1}{2}(\vec{\sigma^{t}} + \vec{\sigma^{p}}), \ \ \
s = 0 (s), 1(t)
\end{eqnarray}
At times we denote the $s=0$, singlet state with ``$s$''
and the $s=1$, triplet state with ``$t$''.

The $\cal{Y}$'s in (\ref{expand})-(\ref{expand2}) are spin--angle
functions and are given in our conventions\cite{sign} as:
\begin{eqnarray}
{\cal Y}^{jm}_{ls}(\hat{k})
	&=& \sum_{m_{s}m_{l}} \langle lm_{l}sm_{s} | jm\rangle
 \,Y_{l}^{m_{l}}(\theta,\phi) \,|s m_{s}\rangle\\
Y_{l}^{m}(\theta,\phi) &=& (-1)^{m} \sqrt{\frac{(2l+1)(l-m)!}{4\pi
(l+m)!}} \, P_{l}^{m}(x\equiv\cos \theta) e^{im\phi}\\
 P_{l}^{m}(x) &=& (1-x^{2})^{m/2} \frac{d^{m} P_{l}(x)}{dx^{m}}
\end{eqnarray}

\noindent To express the Lippmann--Schwinger equation (\ref{LS}) in
the partial wave basis, we substitute (\ref{expand}) and
(\ref{expand2}), the partial--wave expansions of $T$ and $V$, into
(\ref{LS}) and employ the orthogonality of the spin-angle functions,
\begin{equation}
\int {\cal Y}_{ls}^{\dagger jm}(\hat{k})\,
 	{\cal Y}^{jm'}_{l's'}(\hat{k})\, d\hat{k} =
	\delta_{ll'} \delta_{ss'} \delta_{mm'} \label{ortho}
\end{equation}
There results:
\begin{equation}
        T_{l'l}^{j(s's)}(k',k) = V_{l'l}^{j(s's)}(k',k) + \frac{2} {\pi}
\sum_{L\,S}\int_{0}^{\infty}p^2\,dp
\frac{V_{lL}^{j(s'S)}(k',p)T_{Ll'}^{j(Ss)}(p,k)} {E^{+} -
E(p)} \label{LS2}
\end{equation}
In equation (\ref{LS2}), the sum over $L\,S$ is over the orbital
angular momenta and spin states which are allowed to couple for a
fixed value of the total angular momentum $j$.  Accordingly, the
integral equations to solve in the partial wave basis are (we leave
off the $(k',k)$ dependence of the leftmost $T$'s and $V$'s):
\begin{equation}
\left[\begin{array}{c}T^{j(ss)}_{j\,j} \\
T^{j(ts)}_{j\,j} \end{array} \right] =
\left[\begin{array}{c} V^{j(ss)}_{j\,j} \\
V^{j(ts)}_{j\,j} \end{array} \right] +
\int^{\infty}_{0} \! \frac{p^{2} dp}{E^{+} - E(p)}
\left[ \begin{array}{c} V^{j(ss)}_{j\,j}(k'p)\ V^{j(st)}_{j\,j}(k',p) \\
V^{j(ts)}_{j\,j}(k',p)\ V^{j(tt)}_{j\,j}(k',p)
\end{array} \right] \left[ \begin{array}{c} T^{j(ss)}_{j\,j}(p,k) \\
T^{j(ts)}_{j\,j}(p,k) \end{array} \right] 	\label{mix1}
\end{equation}
\begin{equation}
\left[\begin{array}{c}T^{j(tt)}_{j\,j} \\
T^{j(st)}_{j\,j} \end{array} \right] =
\left[\begin{array}{c} V^{j(tt)}_{j\,j} \\
V^{j(st)}_{j\,j} \end{array} \right] +
\int^{\infty}_{0} \!\frac{p^{2} dp}{E^{+} - E(p)}
\left[ \begin{array}{c} V^{j(tt)}_{j\,j}(k',p)\ V^{j(ts)}_{j\,j}(k',p) \\
V^{j(st)}_{j\,j}(k',p)\ V^{j(ss)}_{j\,j}(k',p)
\end{array} \right] \left[ \begin{array}{c} T^{j(tt)}_{j\,j}(p,k) \\
T^{j(st)}_{j\,j}(p,k) \end{array} \right]	\label{mix2}
\end{equation}
\begin{equation}
\left[\begin{array}{c}T^{j(tt)}_{j-1\,j-1} \\
T^{j(tt)}_{j+1\,j-1} \end{array} \right] =
\left[\begin{array}{c} V^{j(tt)}_{j-1\,j-1} \\
V^{j(tt)}_{j+1\,j-1} \end{array} \right] +
\int^{\infty}_{0}\! \frac{p^{2} dp}{E^{+} - E(p)}
\left[ \begin{array}{c} V^{j(tt)}_{j-1\,j-1}(k',p)\ V^{j(tt)}_{j-1\,j+1}(k',p)
\\
V^{j(tt)}_{j+1\,j-1}(k',p)\ V^{j(tt)}_{j+1\,j+1}(k',p)
\end{array} \right] \left[ \begin{array}{c} T^{j(tt)}_{j-1\,j-1}(p,k) \\
T^{j(tt)}_{j+1\,j-1}(p,k) \end{array} \right]	\label{triplet1}
\end{equation}
\begin{equation}
\left[\begin{array}{c}T^{j(tt)}_{j+1\,j+1} \\
T^{j(tt)}_{j-1\,j+1} \end{array} \right] =
\left[\begin{array}{c} V^{j(tt)}_{j+1\,j+1}\\
V^{j(tt)}_{j-1\,j+1} \end{array} \right] +
\int^{\infty}_{0} \!\frac{p^{2} dp}{E^{+} - E(p)}
\left[ \begin{array}{c} V^{j(tt)}_{j+1\,j+1}(k',p)\ V^{j(tt)}_{j+1\,j-1}(k',p)
\\
V^{j(tt)}_{j-1\,j+1}(k',p)\ V^{j(tt)}_{j-1\,j-1}(k',p)
\end{array} \right] \left[ \begin{array}{c} T^{j(tt)}_{j+1\,j+1}(p,k) \\
T^{j(tt)}_{j-1\,j+1}(p,k) \end{array} \right]	\label{triplet2}
\end{equation}

\noindent Equations (\ref{mix1})-(\ref{mix2}) describe spin
singlet-triplet coupling arising from the $V_{f}$ term in the
potential (\ref{V}) which in turn will produce an $f$ amplitude in $T$
(\ref{T}). Equations (\ref{triplet1})-(\ref{triplet2}) describe
coupling within the spin triplet state arising from the tensor force
terms $V_{a-b}, V_{c+d}, V_{c-d}$ in the potential (\ref{V}), it mixes
orbital angular momentum states\cite{haftel}. Since the total angular momentum
$j$
is a conserved quantity, all coupled states have the same $j$
superscript.

\section{Evaluation of $V_{l\,'l}^{j(s's)}$}

We use a two--step procedure to determine the potential matrix
elements $V_{l\,'l}^{j(s's)}(k',k)$ needed in
(\ref{mix1})-(\ref{triplet2}). First we evaluate the potential
(\ref{V}) in the spin basis $|s, m_{s}\rangle$ and then we invert the
angular momentum decomposition of these spin--basis potentials.  The
spin matrix elements required for the $V_{a}-V_{e}$ terms of (\ref{T})
are given in Tables 7.1-7.4 of Goldberger and Watson\cite{gold}. As
expected, the $V_{a}-V_{e}$ terms have vanishing matrix elements
between singlet and triplet states. The new term $V_{f}$ has
nonvanishing matrix elements only between singlet and triplet states,
explicitly
\begin{equation}
   \langle 0, 0 |(\vec{\sigma}^{p}_{n} - \vec{\sigma}^{t}_{n})|1, 1 \rangle \
=  \langle 0, 0 |(\vec{\sigma}^{p}_{n} - \vec{\sigma}^{t}_{n})|1, -1 \rangle \
= - i\sqrt{2} \label{new}
\end{equation}
There is no $\theta$ or $\phi$ dependence in (\ref{new}) because we
adopt the ``Madison Convention'' shown in Figure 1. This convention
takes the $z$-axis as the beam direction $\vec{k}$ ($\phi_{i} =
\theta_{i} = 0$), and places the scattered momentum $\vec{k'}$ in the
$xz$ plane ($\theta_{f} = \theta, \phi_{f} = 0$). Accordingly:

\begin{eqnarray}
           V_{ss}(\vec{k'},\vec{k}) &\equiv&   \langle 0, 0 |V| 0,
0\rangle \nonumber \\
      &=&  V_{a+b}(\vec{k'},\vec{k}) -V_{a-b}(\vec{k'},\vec{k})
-V_{c+d}(\vec{k'},\vec{k}) -V_{c-d}(\vec{k'},\vec{k})\\
 V_{s1}(\vec{k'},\vec{k}) &\equiv&   \langle 0, 0 |V|1, 1\rangle
            = -V_{1s}(\vec{k'},\vec{k}) = V_{s-1}(\vec{k'},\vec{k}) \ = \
               \frac{-i}{\sqrt{2}} V_{f}(\vec{k'},\vec{k}) \label{s1}\\
  V_{00}(\vec{k'},\vec{k}) &\equiv&   \langle 1, 0 |V|1, 0 \rangle \nonumber
\\
 &=&  V_{a+b}(\vec{k'},\vec{k}) + V_{a-b}(\vec{k'},\vec{k})
 + \left(V_{c+d}(\vec{k'},\vec{k}) - V_{c-d}(\vec{k'},\vec{k})\right)\cos\theta
\\
  V_{11}(\vec{k'},\vec{k}) & = & V_{-1-1} = V_{a+b}(\vec{k'},\vec{k})
+V_{c+d}(\vec{k'},\vec{k})\sin^{2}\frac{\theta}{2}
             + V_{c-d}(\vec{k'},\vec{k})\cos^{2}\frac{\theta}{2} \\
  V_{10}(\vec{k'},\vec{k}) & = & -V_{-10} \nonumber \\
 &=& \frac{-i}{\sqrt{2}} V_{e}(\vec{k'},\vec{k})
-\frac{1}{\sqrt{2}}V_{c+d}(\vec{k'},\vec{k})\sin\theta + \frac{1}{\sqrt{2}}
V_{c-d}(\vec{k'},\vec{k}) \sin\theta \\
  V_{01}(\vec{k'},\vec{k}) & = & - V_{0-1} \nonumber \\
	&=& \frac{i}{\sqrt{2}} V_{e}(\vec{k'},\vec{k}) - \frac{1}{\sqrt{2}}
V_{c+d}(\vec{k'},\vec{k}) \sin\theta
           + \frac{1}{\sqrt{2}} V_{c-d}(\vec{k'},\vec{k}) \sin\theta \\
  V_{1,-1}(\vec{k'},\vec{k}) & = & V_{-11} = -V_{a-b}(\vec{k'},\vec{k}) +
V_{c+d}(\vec{k'},\vec{k})\cos^{2}\frac{\theta}{2}
             + V_{c-d}(\vec{k'},\vec{k}) \sin^{2}\frac{\theta}{2}
\end{eqnarray}

Here we have the matrix elements of $V$ in the spin basis. We next
expand these matrix elements in angular momentum states in order to
determine the partial--wave matrix elements $V^{j(s's)}_{l'l}$.  We
take the expansion of the potential in spin--angle functions
(\ref{expand}) and evaluate the matrix element between spin states:
\begin{eqnarray}
     \langle s''' m_{s}'''| V(\vec{k'},\vec{k})|s'' m_{s}''\rangle &=&
  \frac{2}{\pi} \sum_{js'sl'lm_{s}m_{s'}m_{l}m_{l'}} \langle s'''m_{s}'''
  |s'm_{s}'\rangle \,i^{l - l'}\,
\langle l'm_{l}'s'm_{s}'|jm\rangle \nonumber \\
\times \langle jm|lm_{l}sm_{s}
\rangle
& &Y^{m_{l'}}_{l'}(\theta_{f},\phi_{f}) \, V^{js's}_{l'\,l}(k',k)\,
Y^{m_{l}^{\ast}}_{l}(\theta_{i},\phi_{i})\, \langle
sm_{s}|s'',m_{s}''\rangle \label{V3}
\end{eqnarray}
where $(k, \theta_{i}, \phi_{i})$ and $(k', \theta_{f}, \phi_{f})$ are
the spherical coordinates of the initial and final momenta.  The
Clebsch-Gordon coefficients vanish unless $l, l' = j\pm 1, j$ and $m_{j} =
m_{l} + m_{s} = m_{l}' + m_{s}'$.  Parity conservation requires $|l -
l'| = 0, 2$. In the ``Madison Convention'' (Figure 1) the projectile
has no angular momentum in its propagation direction and so $m_{l} =
0$, in which case
\begin{equation}
Y^{m_{l}^{\ast}}_{l}(\theta_{i},\phi_{i})
= Y^{0}_{l}(0, 0) =  \sqrt{\frac{2l + 1}{4\pi}}
\end{equation}

As a sample, we concentrate on the new term $V_{s1}$ (\ref{s1}) which
couples the $|00 \rangle$ singlet state to the $|11 \rangle$ triplet
state. Because $j$ is a constant and $s' = 0$ in the final state, the
total angular momentum $j$ must equal $l'$.  The parity constraint
then requires that $l = l'$.  Because $m_{s}$ and $m_{l}$ are 1 and 0,
respectively, we deduce that $m_{j} = 1$, and $m_{l'} = 1$.  For this
$V_{s1}$ term, the sum in (\ref{V3}) reduces to a simple sum in the
final orbital angular momentum $l'$:

\begin{equation}
\sum_{js'sll'm_{s}m_{s'}m_{l}m_{l'}} \! \! \! \! \! \! ...
\rightarrow \sum_{l} \! ...
\end{equation}
Similarly, when the spin-dependent singlet-singlet potential
$V_{ss}(\vec{k'}, \vec{k})$ and those within the triplet state
$V_{m'm}(\vec{k'}, \vec{k})$ are evaluated, we obtain the desired partial--wave
expansion of $V$ (and of $T$ with the interchange $V\rightarrow T$):
\begin{eqnarray}
V_{s1}(\vec{k'}, \vec{k}) &=& \frac{-\sqrt{2}}{4\pi^{2}}
\sum_{l=1} P_{l}^{1}(x = \cos\theta_{k'k})\, \frac{2l + 1}{\sqrt{l(l + 1)}}
\,V_{l\,l}^{l(st)}(k',k)  \label{vs1} \\
V_{ss}(\vec{k'}, \vec{k}) &=& \frac{1}{2\pi^{2}} \sum_{l=0} P_{l}(x)
         (2l + 1) \, V_{l\,l}^{l(ss)}(k',k) \label{vss} \\
V_{11}(\vec{k'}, \vec{k}) &=& \frac{1}{4\pi^{2}}
\sum_{l=0}P_{l}(x)\left\{(l+2)V_{l\,l}
       ^{l+1(tt)}(k',k) - \sqrt{(l+1)(l+2)} \: V_{l\,
l+2}^{l+1(tt)}(k',k) \right. \label{v11}\\
& & \left.
+ (2l+1) V_{l\,l}^{l(tt)}(k',k)
 + (l-1) V_{l\,l}^{l-1(tt)}(k',k)  -
       \sqrt{(l-1)l}\, V_{l \, l-2}^{l-1(tt)}(k',k)  \right\} \nonumber   \\
V_{00}(\vec{k'}, \vec{k}) &=& \frac{1}{2\pi^{2}} \sum_{l=0}
P_{l}(x)\left\{(l+1)
       V_{l\,l}^{l+1(tt)}(k',k) + l V_{l\,l}^{l-1(tt)}(k',k)
       \right.\nonumber\\ & &\left.
       + \sqrt{(l+1)(l+2)}\, V_{l\, l+2}^{l+1(tt)}(k',k) + \sqrt{(l-1)l}\,
       V_{l\,l-2}^{l-1(tt)}(k',k) \right\} \\
V_{10}(\vec{k'}, \vec{k})
&=& \frac{\sqrt{2}}{4\pi^{2}} \sum_{l=1} P_{l}^{1}(x) \left\{-
       V_{l\,l}^{l-1(tt)}(k',k)\! + V_{l\,l}^{l+1(tt)}(k',k)
\right.\nonumber\\
  & &\left.+ \sqrt{\frac{l+2}{l+1}}
       V_{l\,l+2}^{l+1(tt)}(k',k)
- \!\sqrt{\frac{l-1}{l}}V_{l\,l-2}^{l-1(tt)}(k',k)
       \right\} \\
V_{01}(\vec{k'}, \vec{k}) &=& \frac{\sqrt{2}}{4\pi^{2}}
\sum_{l=1} P_{l}^{1}(x)\left\{
       -\frac{l+2}{l+1} V_{l \,l}^{l+1(tt)}(k',k)
       + \frac{2l+1}{l(l+1)}
       V_{l \,l}^{l(tt)}(k',k)  \right.
       \\ & &\left. + \frac{l-1}{l} V_{l \,l}^{l-1(tt)}(k',k)
	+ \sqrt{\frac{l+2}{l+1}} V_{l\,l+2}^{l+1(tt)}(k',k) -
       \sqrt{\frac{l-1}{l}} V_{l\,l-2}^{l-1(tt)}(k',k) \right\} \nonumber\\
V_{1\,-1}(\vec{k'}, \vec{k}) &=& \frac{1}{4\pi^{2}}
	\sum_{l=2} P_{l}^{2}(x) \left\{
        \frac{1}{l+1} V_{l\,l}^{l+1(tt)}(k',k)
        - \frac{1}{\sqrt{(l+1)(l+2)}} V_{l\,l+2}^{l+1(tt)}(k',k)
\right.\nonumber\\
& & \left. - \frac{2l+1}{l(l+1)}
        V_{l\,l}^{l(tt)} (k',k)
 + \frac{1}{l} V_{l\,l}^{l-1(tt)}(k',k)  -
        \frac{1}{\sqrt{l(l-1)}} V_{l\,l-2}^{l-1(tt)}(k',k) \right\}
	\label{v1-1}
\end{eqnarray}
Note that the sum is actually over the orbital angular momentum $l'$
of the final state, but for notational simplicity we have changed the
label $l'$ to $l$.  Furthermore, note that the organization in
(\ref{vs1})-(\ref{v1-1}) combines matrix elements which multiply the
same Legendre polynomial even though the matrix elements may
correspond to different $j$ values. The Lippmann--Schwinger equations
(\ref{mix1})-(\ref{triplet2}) of course only couple states with the
same $j$.

We invert equations (\ref{vs1})-(\ref{v1-1}) for the partial wave
potentials $V^{j(s's)}_{l'l}(k',k)$ by projections based on the
orthogonality of the associated Legendre polynomials.
Specifically, we multiply the equation for each $V_{m'm}$
by $P_{l}^{|m'-m|}$, and evaluate numerically the integral
\begin{equation}
	I_{m'm}(k',k) = \int_{-1}^{1}dx\, V_{m'm}(\vec{k'},\vec{k})
	\, P_{l}^{|m'-m|}(\cos\theta_{k'k}) \label{I}
\end{equation}
For (\ref{vs1}) and (\ref{vss}) the inversion is simple because only
one $V^{j(s's)}_{l'l}$ is involved:
\begin{eqnarray}
 V^{l(st)}_{l\,l}(k',k)
&=& V^{l(ts)}_{l\,l}(k',k) = \frac{-\sqrt{2}\pi^{2}}{\sqrt{l(l + 1)}}
	I_{s1}(k',k) \label{project1}\\
V^{l(ss)}_{l\,l}(k',k) &=&\pi^{2} I_{ss}(k',k) \label{project2}
\end{eqnarray}
The equations (\ref{v11})-(\ref{v1-1}) contain $V^{j(tt)}_{l'l}$'s
intermixed for differing $j$ and $l$ values, and so the projection
results in five coupled equations in five unknowns:
\begin{equation}
\vec{I} = B \vec{V}
\end{equation}
\begin{equation}
\left[ \begin{array}{c} I_{11}(k',k)\\ I_{00}(k',k)\\
I_{10}(k',k) \\ I_{01}(k',k) \\ I_{1-1}(k',k)
\end{array} \right] = \left[ \begin{array}{c} \\  \\ B^{m'm}_{l'l}\\
\\ \mbox{ }
\end{array} \right] \left[ \begin{array}{c}
V_{l\,l}^{l+1(tt)}(\vec{k'},\vec{k}) \\
V_{l\,l}^{l(tt)}(\vec{k'},\vec{k}) \\
V_{l\,l}^{l-1(tt)}(\vec{k'},\vec{k}) \\
V_{l\, l+2}^{l+1(tt)}(\vec{k'},\vec{k}) \\
V_{l \,l-2}^{l-1(tt)}(\vec{k'},\vec{k})
\end{array} \right] \label{I2}
\end{equation}
where $[B^{m'm}_{l'l}]$ is the matrix of coefficients multiplying the
$V$'s in (\ref{v11})-(\ref{v1-1}). In our application\cite{paez}, the matrix
equation (\ref{I2}) is solved by numerical inversion
\begin{equation}
\vec{V} = B^{-1} \vec{I}
\end{equation}
and then the whole procedure is checked by recombining the potential
according to (\ref{vs1})-(\ref{v1-1}).

Once the $V^{j(s's)}_{l'l}$'s are known, the coupled integral
equations (\ref{mix1})-(\ref{triplet2}) are solved numerically for the
$T^{j(s's)}_{l'l}$'s using a modification of the Haftel--Tabakin
technique\cite{haftel}.  Once the $T^{j(s's)}_{l'l}$'s are known, the
matrix elements in the spin basis $\langle s' m_{s'}|T|s m_{s}
\rangle$ are computed via the generalization of equations
(\ref{vs1})-(\ref{v1-1}) in which the $V$'s are replaced by $T$'s.
Finally, once the $\langle s' m_{s'}|T|s m_{s}
\rangle$ are known, the $a-f$ amplitudes of (\ref{T})
are obtained\cite{laf}:
\begin{eqnarray}
a(\vec{k'},\vec{k}) &=& \frac{1}{2} \left(T_{11}(\vec{k'},\vec{k}) +
T_{00}(\vec{k'},\vec{k}) - T_{1-1}(\vec{k'},\vec{k})\right)\\
b(\vec{k'},\vec{k}) &=&
\frac{1}{2}\left(T_{11}(\vec{k'},\vec{k})+T_{ss}(\vec{k'},\vec{k})+T_{1-1}(\vec{k'},\vec{k})\right) \\
c(\vec{k'},\vec{k}) &=&
\frac{1}{2}\left(T_{11}(\vec{k'},\vec{k})-T_{ss}(\vec{k'},\vec{k})+T_{1-1}(\vec{k'},\vec{k})\right)\\
d(\vec{k'},\vec{k}) &=&
\frac{1}{2}\left(T_{00}(\vec{k'},\vec{k})+T_{1-1}(\vec{k'},\vec{k})-T_{11}(\vec{k'},\vec{k})\right)/2\cos\theta_{k'k}\\
	&=& - \left( T_{10}(\vec{k'},\vec{k}) +
T_{01}(\vec{k'},\vec{k}) \right) / \sqrt{2} \sin\theta_{k'k}\\
e(\vec{k'},\vec{k})
&=&
\frac{i}{\sqrt{2}}\left(T_{10}(\vec{k'},\vec{k})-T_{01}(\vec{k'},\vec{k})\right)\\
f(\vec{k'},\vec{k}) &=& i \sqrt{2} T_{s1}(\vec{k'},\vec{k})
\end{eqnarray}
All spin observables are then calculated from $a-f$ using the
relations found in La France and Winternitz\cite{laf}.

\section{RELATION TO PHASE SHIFTS}

The on-energy-shell $T$ matrix elements in the partial wave basis
$T^{j(s's)}_{l'l}(k_{0},k_{0})$ can be related to phase shifts, a
convenient phenomenological parameterization of the scattering data.
This is particularly useful when the present formalism is applied to
the two nucleon problem because there are tables of NN phase shifts.
The relations to the bar phases are\cite{ger,stapp,haftel}:
\begin{eqnarray}
 -2i\rho T^{j(ss)}_{j\,j}(k_{0},k_{0}) \ \ &=& \ \
\cos2\bar{\gamma}_{l} e^{ 2i\bar{\delta_{j}}} - 1 \\
 -2i\rho T^{j(tt)}_{j\,j}(k_{0},k_{0}) \ \ &=& \ \ \cos2\bar{\gamma}_{l}
e^{2i\bar{\delta}_{jj}} - 1\\
 -2i\rho T^{j \: (tt)}_{j\pm1\,j\pm1}(k_{0},k_{0}) \ \ &=& \ \
\cos2\bar{\epsilon_{j}}
 e^{ 2i\bar{\delta}_{j\pm1 \: j}} - 1 \\
 -2i\rho T^{j\: (tt)}_{j\pm 1\:j \mp1}(k_{0},k_{0}) \ \ &=& \ \ -i \sin
2\bar{\epsilon_{j}}
 \: e^{2i(\bar{\delta}_{j-1\:j} + \bar{\delta}_{j+1\:j})} \\
 -2i\rho T^{j(ts)}_{j\,j}(k_{0},k_{0}) \ \ &=& \ \ -i\sin 2\bar{\gamma}_{l} \:
e^{ i(\bar{\delta}_{j} +
 \bar{\delta}_{jj})} \\
\rho \ \ &=& \ \ 2k_{0} \frac{E_{p}(k_{0}) E_{t}(k_{0})}
{E_{p}(k_{0})+E_{t}(k_{0})}
\end{eqnarray}
The parameter $\bar{\gamma}_{l} $ is the mixing angle between the
$|0,0\rangle$ singlet and $|1,1\rangle$ triplet state, and the
parameter $\bar{\epsilon}_{l}$ is the mixing angle between the $l$ and
$l+2$ triplet states.  In Table I we give the connection to the
$\alpha$ notation of Stapp\cite{stapp} as well as that to the
$N_{spin}$ notation used in our computer code {\em Lpotp2}.

\section{SUMMARY}

We have extended the partial wave analysis of the Schr\"odinger
equation describing the interaction of two spin $\frac{1}{2}$
particles to the case where a potential couples the spin singlet and
triplet states as well as coupling within the triplet state. In
particular, we have concentrated on the scattering configuration
described by the Lippmann--Schwinger equation although the same
formalism can be used for bound states.  While a previous formalism
was appropriate to the nucleon-nucleon problem, extensions are
necessary when the two Fermions are not identical. We are now applying
the new formalism to describe polarized proton scattering from polarized
$^{3}He$ and $^{13}C$ nuclei\cite{me}. Our formalism should also be useful in
describing the interaction between identical hadrons when isospin
symmetry is violated. Furthermore, the formalism may find some
applicability in the electron-atom interaction in atoms such as
calcium where there is strong singlet-triplet mixing.

\acknowledgments

We wish to thank Lanny Ray, Sid Coon, Otto H\"auser, and Charlie Drake
for stimulating and illuminating discussions.  We also wish to
gratefully acknowledge support from the U.S. Department of Energy
under Grant \ DE-FG06-86ER40283, and the people at the National
Institute for Nuclear Theory, Seattle for their hospitality during
part of this work.


\figure{ The coordinate system used to describe the scattering of momentum
$\vec{k}$ into $\vec{k'}$ in the Madison convention. The incident
momentum is along the $z$ axis and the final momentum is in the $xz$
plane. Note that $\theta'$ is the same as the $\theta_{k'k}$ in
equations (\ref{vs1})-(\ref{v1-1}).\label{fig1}}

\widetext
\begin{table}
\caption{Notations for spin $\frac{1}{2} \times
\frac{1}{2}$ amplitudes ($T^{0(tt)}_{0\,0}= T^{0(tt)}_{-1\,-1}=
T^{0(tt)}_{1\,-1}= T^{0(tt)}_{-1\,1}=0$).}
\begin{tabular}{l|cccccccc}
 $ T^{j(s's)}_{l'\,l} $ & $ T^{j(ss)}_{j\,j} $ & $ T^{j(ts)}_{j\,j} $
& $ T^{j(tt)}_{j-1\,j-1} $ & $ T^{j(tt)}_{j+1\,j-1} $ & $
T^{j(tt)}_{j+1\,j+1} $ & $ T^{j(tt)}_{j-1\,j+1} $ & $ T^{j(tt)}_{j\,j}
$ & $ T^{j(st)}_{j\,j}$\\ Spin & 0 & $1 \leftarrow 0$ & 1 & 1 & 1 & 1
& 1 & $0 \leftarrow 1$\\ $\Delta l$ & 0& 0& 0& 2& 0& -2& 0&0\\ Stapp &
$ \alpha_{l} $ & $ - $ & $ \alpha_{l,l+1} $ & $ \alpha^{l-1} $& $
\alpha_{l,l-1} $ & $ \alpha^{l+1} $ & $ \alpha_{ll} $ & $ -$ \\
$N_{spin}$& 1 & 2 &3&4&5& 6 & 7&8

\end{tabular}
\label{table1}
\end{table}


\begin{references}
\bibitem{wave}
Note although neither $T$ nor $V$ is a wave, we refer to their angular
momentum decomposition as a ``partial--wave'' expansion. Note too that in
some previous work, $V^{l(ss)}_{ll}$ was called $V^{l}_{l}$ for short.
\bibitem{sign}
This is the sign convention of Messiah and of Goldberger and Watson. It
differs by $(-1)^m$ from Abrahmowitz and Stegun, Jackson, and
Ref.\cite{god}.
\end{references}
\end{document}